# INVESTIGATING THE INFLUENCE OF PARTICLE SIZE AND SHAPE ON FROTH FLOTATION BASED BENEFICIATION OF LITHIUM-RICH MINERALS IN SLAGS

*Franziska Strube[1], Thomas Wilhelm[2], Johanna Sygusch[1], Bradley M. Guy[1], Orkun Furat[2], Volker Schmidt[2], Martin Rudolph[1]*

1 Helmholtz-Zentrum Dresden-Rossendorf, Helmholtz Institute Freiberg for Resource Technology, Chemnitzer Straße 40, 09599 Freiberg, Germany

2 Ulm University, Institute of Stochastics, Helmholtzstraße 18, 89069 Ulm, Germany

## ABSTRACT

The demand for lithium, as well as other critical resources, needed for electrochemical energy storage is expected to grow significantly in the future. Slags obtained from pyrometallurgical recycling represent a promising resource of valuable materials, among them lithium and rare earth elements found in artificial minerals particulate phases. This study investigates the flotation separation of engineered artificial minerals (EnAMs) in slags, such as lithium aluminate and gehlenite as valuable and gangue phases, respectively. Flotation experiments are carried out in a Partridge-Smith cell using oleic acid (OA) as a benchmark surfactant. Particle characterization is performed using SEM-based Mineral Liberation Analysis (MLA), which provides particle discrete information. From this information, bivariate Tromp functions based on non-parametric kernel density estimation are computed to characterize the separation behavior with respect to particle descriptors. This approach enables investigating the influence of particle size and shape on separation behavior of EnAMs. Furthermore, these results allow for the optimization of flotation experiments for enriching Li-bearing EnAMs.

## INTRODUCTION

In the past years, the interest in lithium, particularly as the main component of lithium-ion batteries (LIBs), has increased dramatically. As the global society tries to integrate renewable energy to decrease carbon emission for limiting climate change, the demand for LIBs has doubled in the last five years. (Tabelin et al. 2021) It is the most important renewable energy storage system and especially applied for electric-base vehicles. Alternative battery storage systems, which are not using lithium are still in development, but they are far from having the same capacity like LIBs and are not soon realized in industry (Tabelin et al. 2021). Because of its outstanding importance for modern society, lithium was declared a critical element and was even referred to as "white" gold (Tarascon 2010).

In nature, lithium is primarily found in salt brines located near volcanic centres, with significant deposits also occurring in spodumene-bearing pegmatites. Smaller quantities are also found in clays, sediment deposits and geothermal brines. It is estimated that the global demand for lithium will increase significantly in the next years. Therefore, the development of efficient recycling and beneficiation routes for lithium is crucial. The key methods currently available for recycling of LIBs include pyrometallurgy and hydrometallurgy or a hybrid of both (Tabelin et al. 2021). Pyrometallurgical recycling has the advantage to be applicable to many waste streams and already existing plants can be used, e.g. for copper and nickel recycling. Due to its ignoble character, lithium is concentrated in the slag beside the noble main components with high melting temperatures, such as cobalt and nickel.
Usually, LIBs contain between 2 to 15 % lithium depending on the LIB technology and thus can contain even more lithium than natural resources (Maarten Quix et al. 2017). That's why the further processing of lithium-bearing slags becomes more attractive and offers a novel way to recover lithium from the waste stream. A goal is to enrich lithium into artificial minerals during slagging and to adjust cooling rates to gain larger crystals, which facilitates the processing. The most prominent engineered artificial mineral (EnAM) is lithium aluminate ($LiAlO_2$), which contains the highest lithium molar ratio. It is embedded into the gangue phase gehlenite ($Ca_2Al(AlSiO_7)$) which is a sorosilicate (Elwert et al. 2012).

Both, the natural Li-bearing spodumene as well as the slag with Li-bearing EnAMs, need to undergo upgrading processes until they can be further transferred into $Li_2CO_3$ for industrial applications. First, common mineral processing operations like crushing and milling are used, followed by magnetic separation, density separation and flotation to produce concentrates containing 4-8 % $Li_2O$ (Tabelin et al. 2021). In order to improve the separation stage of flotation, more efficient reagent regimes, especially for the flotation of spodumene, need to be developed.

Usually anionic collectors are used for spodumene flotation, like the benchmark collectors sodium oleate (Filippov et al. 2019) and oleic acid (OA) (Tadesse et al. 2019; Yu et al. 2015) because they offer higher selectivity than cationic collectors. It was found that sodium oleate not only works as a collector for spodumene but also for $LiAlO_2$ (Qiu et al. 2021). Recently, research was done on both sources with new and mixed collector systems of cationic and anionic collectors for spodumene (Tian et al. 2017, Acker et al. 2023).
In this study OA was chosen as a collector for the flotation of lithium bearing EnAMs, in order to establish a benchmark case for this type of slag.

Although, the flotation is strongly determined by the differences in particle wettability, other particle properties, such as shape, size or liberation, also play a crucial role for the separation behaviour. Recoveries of particles that are either too fine or too coarse are generally low, due to slow flotation kinetics of fines and less stable particle-bubble aggregates for coarse particles. (Schubert 1996; Dai et al. 2000; Gontijo et al. 2007)
Investigations with respect to the particle shape are a bit more diverse in their results, as the influence seems to depend on the particle system itself, especially on the size class studied, and also on the flotation apparatus used, i.e., if the flotation was carried out in a mechanical cell, flotation column or micro flotation etc. (Koh et al. 2009; Xie et al. 2017; Sygusch 2023; Hassas et al. 2016).
With all these properties adding to the complexity of the separation, a multidimensional evaluation is needed in order to understand the interplay of said descriptors, which can then be used to improve the beneficiation of the Li-bearing EnAMs. (Wilhelm et al. 2023) demonstrated that bivariate Tromp functions can be computed from scanning electron microscopy based image data of the input and the output streams. With this method, the influence of the particle shape (aspect ratio) and size (area-equivalent diameter) on the separation behaviour of $LiAlO_2$ via flotation is investigated.

## METHODS AND MATERIALS

### Materials

The slag which was provided by RWTH Aachen University (IME Process Metallurgy and Metal Recycling, RWTH Aachen, Germany) is a mock slag close to the composition of real LIBs slags $Li_2O$-$CaO$-$SiO_2$-$Al_2O_3$-$MgO$-$MnO_x$ (Rachmawatil et al. 2024, Wittkowski et al. 2021). The particular initiating composition was $Li_2O$ 8.5 mol%, $Al_2O_3$ 45 mol%, $SiO_2$ 19 mol%, $CaO$ 17 mol%, 10 mol% $MnO$ and the applied cooling rate was 25 °C/h with a batch size of 120 kg.

For the sample preparation 80 kg of slag were crushed in a jaw crusher in 30 cm blocks at UVR-FIA GmbH, Freiberg, Germany. The material was crushed three times to a particle size smaller than 10 mm. Afterwards the material was given into a sieve ball mill and milled to a particle size smaller than 100 µm. The result is a red brown powder, as seen on Figure 1.

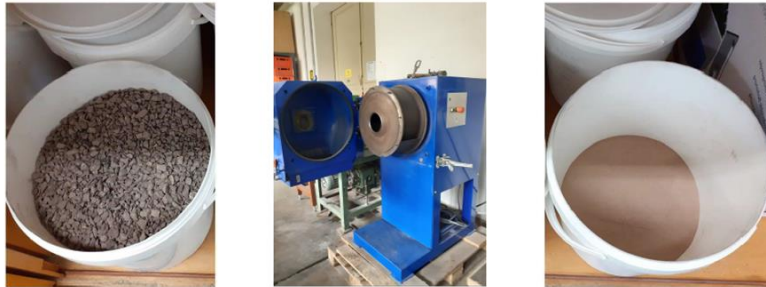

**Figure 1:(left) slag material after crushing < 10 mm, (middle) used sieve ball mill with a sieve size of 100 μm, (right) slag powder after milling <100 μm**

Solutions of 0.01 g/ml of oleic acid as collector (Sigma-Aldrich, St. Louis, MO, USA) and methyl isobutyl carbinol, MIBC, as frother (Sigma-Aldrich, St. Louis, MO, USA) were prepared in ethanol (Technical grade, Carl Roth GmbH + Co. KG, Karlsruhe, Germany).

## Methods

### Batch flotation process

Prior to the froth flotation processes 5 kg slag powder was split with a rotary splitter into roughly 600 g and further into 25 g samples. The batch size froth flotation was performed in a Partridge-Smith cell using the Dynamic Foam Analyser 100 from KRÜSS GmbH, Germany, with a volume of 200 ml. (Figure 2)

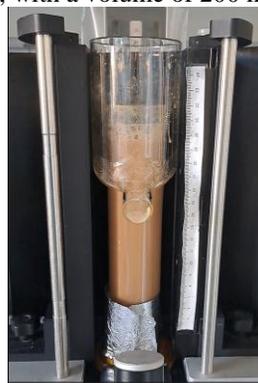

**Figure 2: Partridge-Smith cell using the Dynamic Foam Analyser 100 from KRÜSS GmbH, Germany, with a volume of 200 ml.**

Flotation was carried out with a mass fraction of 11 % solids to 200 ml tap water. For conditioning, the slag was suspended in 100 ml tab water and oleic acid was added in the concentration of 500 g/t, 750 g/t and 1000 g/t with a conditioning time of 5 minutes. Afterwards, 70 g/t of MIBC was added. The batch sized froth flotation was carried out with an air flow rate of 0.8 l/min. Three concentrates (C) were taken after 30 (C1), 60 seconds (C2) and after the flotation stopped (C3). The remaining suspension in the flotation cell was balanced as tailing. After flotation, the concentrates and tailing were filtered and dried at 60°C for approximately 48 hours. The samples were weighed and deagglomerated with a ball shaking machine (MIXER MILL MM 400, Retsch GmbH). Small sample amounts were deagglomerated with a spatula technique to avoid material loss. For the MLA sample preparation an amount of 1-2 g of sample was embedded in epoxy resin, while the rest was analysed via laser diffraction and dynamic image analyses. Every reagent setting was repeated three times. The respective concentrates and tailings of the three tests of one condition (500 g/t, 750 g/t, 1000 g/t) were mixed together to obtain an average sample with sufficient mass for analysis.

*Mineral Liberation Analysis (MLA)*

Twelve samples containing the concentrates and tailings were analysed with MLA. All samples were carbon coated using a Leica MED 020 vacuum evaporator to ensure conductivity of the sample surface. The MLA is an FEI Quanta 650F scanning electron microscope equipped with two Bruker Quantax X-Flash 5030 EDX detectors and FEI's MLA Suite 3.1.4 for automated data acquisition. The electron beam accelerating voltage was 25 kV. The MLA measurement method of choice was GXMAP and the X-ray step size spacing was set 6 µm. The resolution was 700 µm and the scan speed 16.

*Stochastic modeling of particle descriptor vectors for the computation of bivariate Tromp functions*

A particle system, observed by image measurements and extracted by particle-based segmentation, can be effectively described using descriptor vectors. Particularly, in this study, the application of MLA images offers planar sections of a three-dimensional particle system. Here, each particle is represented by a 2-dimensional descriptor vector derived from the particle-wise segmentation of 2D images, containing the particle's area-equivalent diameter ($d_A$) and aspect ratio ($\varphi$). These descriptors are adapted from Wilhelm et al. (2023). In order to evaluate the separation behavior of particle systems resulting from froth flotation tests, the entirety of particle descriptor vectors associated with particles of the feed material (F) and of the complete concentrate (C) is modeled by mass-weighted bivariate probability densities $f_m^F$ and $f_m^C$, respectively. In the present study, we only have image measurements of each of the three concentrates (C1, C2 and C3) and tailings (T). Thus, we computed the number-weighted probability density $f_n^C$ and $f_n^T$ of descriptor vectors associated with particles in the concentrate and tailings, respectively, before computing $f_m^F$. More precisely, the probability densities $f_n^{C1}$, $f_n^{C2}$, $f_n^{C3}$ and $f_n^T$ were computed from image measurements using kernel density estimators, as outlined by Scott (2015). Then, as explained in Wilhelm et al. (2023), these probability densities were transformed to mass-weighted probability densities $f_m^{C1}, f_m^{C2}, f_m^{C3}, f_m^T$ and the probability density $f_m^C$ was computed. Afterwards, the probability density $f_m^F$ is determined as a convex combination of $f_m^C$ and $f_m^T$, as discussed in Buchmann et al. (2018) and Wilhelm et al. (2023). This enables the computation of the bivariate Tromp function $T$ as the ratio of $f_m^C$ and $f_m^F$ multiplied with the mass ratio of particles observed in concentrate and feed, where the value $T(d_A, \varphi)$ indicates the probability that a particle with descriptor vector $(d_A, \varphi)$ is separated into the concentrate, as explained in Wilhelm et al. (2023). In the present paper, we determine $T$ for three different froth flotation tests involving slag particles.

**RESULTS AND DISCUSSION**

**Composition and characteristics of slag particles**

In total the provided slag shows four main phases including the most prominent phases $LiAlO_2$ (7.2 wt%) and gehlenite (32.5 wt%). The elemental composition of $LiAlO_2$ contains an additional content of Si (1-5 wt%) and gehlenite shows an incorporated manganese content of 2-8 wt%. Furthermore, the main components are the Li-bearing lithium aluminium silicate eucryptite (13.5 wt%) and aluminium manganese spinel types (29.5 wt%) with varying manganese content. Particularly the varying spinel types could be already observed from Wittowski et al (Wittkowski et al. 2021). In the present slag one major AlMn-spinel type occurs with 3-12 wt% Mn and another minor Mn-rich AlMn-spinel with 27-47 wt% Mn. Furthermore, a finely grained mixed phase exists with the elemental composition of Mn (17-28 wt%), Si (20-28 wt%), Al (13-18 wt%) and Ca (0-3 wt%). The included mixing phases can be mainly eucryptite, AlMn containing spinel-types, $LiMnSiO_4$ as well as inclusions of gehlenite. (Table 4 ).

**Table 4: Mineral phase composition of the present slag. Shown is the weight fraction of each phase determined from the area % from MLA information and the sample weight. Besides the four main phases and a mixed phase, other components with 1.6 wt% including the phases like glaucochroite (CaMnSiO$_4$) and LiMnSiO$_4$, which are also known phases during slag formation in the present slag system Li$_2$O-CaO-SiO$_2$-Al$_2$O$_3$-MnO$_x$ (Wittkowski et al. 2021).**

| name | lithium aluminate | eucryptite | gehlenite | AlMn-spinel type | mixed phase | others |
|---|---|---|---|---|---|---|
| formula/ consisting of elements | LiAlO$_2$ | LiAl[SiO$_4$] | Ca$_2$Al(AlSiO$_7$) | Al,Mn - oxide | Si, Mn, Al, Ca - oxide | varying (Si, Mn, Al, Ca, Ti, Cr) |
| fraction in wt % | 7.2 | 13.5 | 32.6 | 29.5 | 15.6 | 1.6 |

The most abundant phases in the Li-slag is gehlenite with 32.6 wt% and. AlMn-spinel type with 29.5 wt %. The evaluation of the MLA images of slag particles show that grains of LiAlO$_2$ are finely distributed in the gangue phase gehlenite. In total roughly 73 % of LiAlO$_2$ grains in slag particles are between 7 µm and 53 µm of the grain size, hence a full liberation of each phase into single particles can be challenging. (Figure 3) This is a possible reason that the MLA images show only 3-4 % of particles contain 80-100 area% of LiAlO$_2$, which can be declared as liberated. In total only 0.07 % of LiAlO$_2$ is fully liberated into particles with100 area% LiAlO$_2$, which represent in total 36 % of the entire LiAlO$_2$ area in the sample. Roughly 50 % of the LiAlO$_2$ area is covering 30-80 area% of the particles and are declared as middlings. Furthermore,17 % LiAlO$_2$ appeares as locked target phase, which covers only 0-30 area% of the particle area. 70 % of the measured particles don't contain any LiAlO$_2$. Whereas, gehlenite is liberated to 14-15 % (17 % of particles contain no gehlenite, 45 % locked, 26 % as middlings) and AlMn-spinel type to 8 % (35 % contain no AlMn-spinel, 23 % locked, and 35 % as middlings) into particles. Since these are the most abundant phases their liberation is more likely.

LiAlO$_2$ occurs as idiomorphic grains and is mainly accompanied in middlings with gehlenite, second most common with the EnAM eucryptite and third most common with AlMn-spinel type. When a major part of particles containing middlings of LiAlO$_2$ accompanied with gehlenite can influence the flotation as the aim is to separate the valuable LiAlO$_2$ from gehlenite.

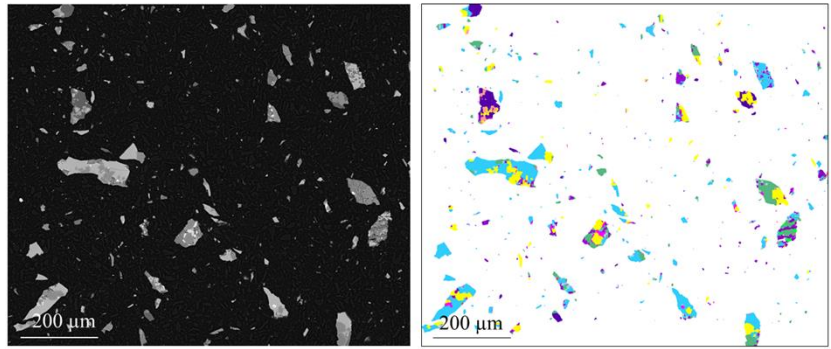

**Figure 3: Shows representative images for (left) BSE backscatter image of particles embedded in graphite epoxy resin, (right) processed MLA image (e.g. violet = LiAlO2, blue = gehlenite, yellow = AlMn-spinel, pink = Mnrich AlMn-spinel, green = mixed phase).**

**Froth flotation**

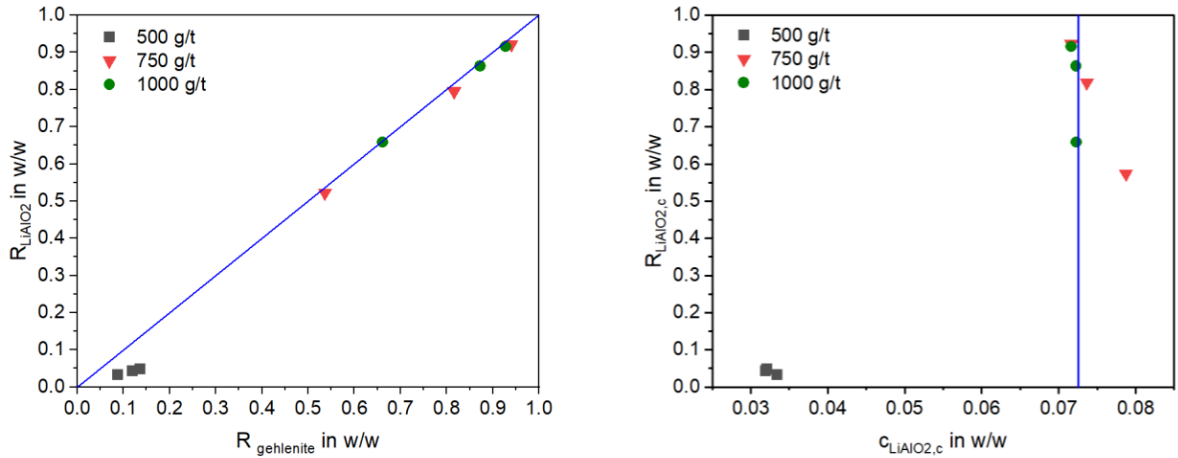

Figure 4: Displays the results (left) Fuerstenau upgrading diagram showing selectivity as a result for the slag batch flotation tests with 13 wt% slag, variation of added collector OA with 500 g/t, 750 g/t and 1000 g/t and frother concentration of 70 g/t MIBC at pH 10.5. The blue solid line represents the splitting curve where no enrichment occurs. (right) Halbich upgrading curve plotting the recovery of LiAlO$_2$ in the concentrates against the grade of LiAlO$_2$ in the concentrates. The blue line shows the feed grade $c_{LiAlO2,c}$.

At a concentration of 500 g/t OA gehlenite is enriched instead of LiAlO$_2$. This matches the results seen in the Halbich-plot, as the grade of LiAlO$_2$ in the concentrate decreases, compared to the feed grade. The absolute mass pull at 500 g/t is relatively low in comparison to the other collector concentrations. From Figure 5 it can be observed that the bubbles of the froth are rather loaded with particles and the froth appears to be drier, compared to that of higher concentrations. However, although the absolute water pull is rather low, the concentration of 500 g/t OA still promotes the recovery of fine particles into the concentrates with 90 % being smaller than 61 µm (Figure 6, right).

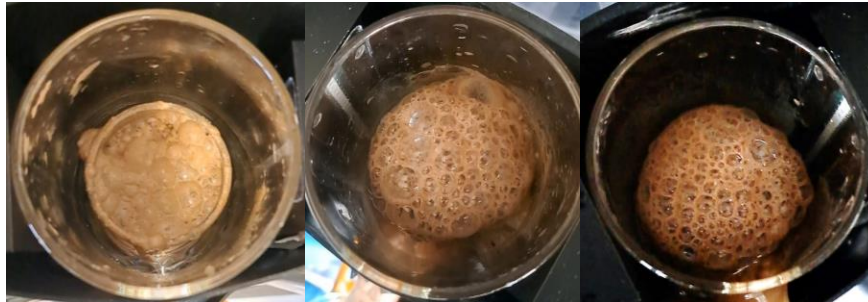

Figure 5: Froth at flotation time of 30 s until the first concentrate was taken, (left) at a collector concentration of 500 g/t OA, (middle) 750 g/t OA and (right) 1000 g/t.

As the concentration of OA increases, the absolute mass and water pull increase as well. At a concentration of 1000 g/t $_{OA}$ almost 90 wt% of the entire feed mass is found in the concentrates (Table 1), resulting in a similar grade of LiAlO$_2$ (Figure 4, right) and a similar size distribution, compared to the feed (Figure 6, right). The same tendency can be observed for the tests using 750 g/t OA. Although the selectivity for gehlenite, as seen in the Fuerstenau-plots for the tests using 500 g/t, is reduced if higher OA amounts are used, there is still no selectivity for LiAlO$_2$, as all data points lie on the blue line. This indicates, that the recovery of particles with higher OA amounts is rather unselective, since almost all particles of the feed report to the concentrate, regardless of their composition.

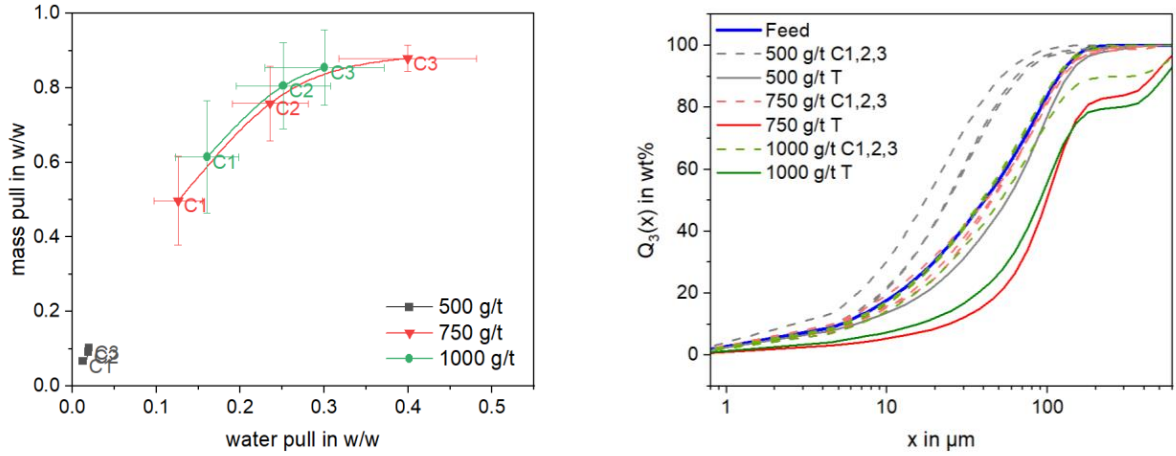

**Figure 6: (left) mass pull plotted against water pull, (right) sum distribution of particle sizes in feed, concentrates C1, C2, C3 and tailing with 500, 750 and 1000 g/t OA addition in froth flotation.**

The probability densities $f_n^C$, $f_n^T$ and $f_n^F$ of descriptor vectors were computed associating with particles of three different particle systems: all particles, particles exclusively containing $LiAlO_2$ or gehlenite. In the analysis of the latter two particle systems, particles that contain both $LiAlO_2$ and gehlenite were excluded. By conducting these analyses, a varied separation behavior among these different particle types was investigated. From the bivariate densities, the feed, concentrate and tailings of the froth flotation tests were characterized using aggregated descriptors. i.e., the mean area-equivalent diameter ($\overline{d_A}$) as listed in Tables 2 and the mean aspect ratio ($\overline{\varphi}$). A noteworthy observation is that particles containing $LiAlO_2$ tend to be smaller than other particles. Across all froth flotation tests smaller particles are more likely to be separated into the concentrate. With an increase in OA concentration, coarse particles are more likely to be enriched in the concentrate. However, when comparing concentrations of 1000 g/t and 750 g/t, the $\overline{d_A}$ of particles in the concentrate decreases again. Moreover, $\overline{\varphi}$ of particles remains consistent 0.59-0.64 across the feed, concentrate, or tailings, and the varying OA concentrations do not affect the separation behavior.

**Table 1: Mean area-equivalent diameter ($\overline{d_A}$ in $\mu m$) associated with all particles, of $LiAlO_2$ particles and gehlenite particles in feed, concentrate and tailings for froth flotation tests with 500 g/t, 750 g/t and 1000 g/t OA.**

| | oleic acid concentration | Feed | concentrate | tailings |
|---|---|---|---|---|
| all particles | 500 g/t | 10.05 | 8.75 | 10.67 |
| | 750 g/t | 10.38 | 10.37 | 17.39 |
| | 1000 g/t | 9.87 | 9.75 | 14.23 |

| | oleic acid concentration | Feed | concentrate | tailings |
|---|---|---|---|---|
| $LiAlO_2$ | 500 g/t | 8.44 | 8.66 | 6.92 |
| | 750 g/t | 8.65 | 8.65 | 12.71 |
| | 1000 g/t | 8.47 | 8.38 | 10.97 |

| | oleic acid concentration | feed | concentrate | tailings |
|---|---|---|---|---|
| gehlenite | 500 g/t | 11.28 | 9.43 | 11.28 |
| | 750 g/t | 11.94 | 11.88 | 21.25 |
| | 1000 g/t | 11.39 | 11.25 | 16.61 |

## Tromp function for 2-dimensional particle descriptor vectors

Bivariate Tromp functions were computed for all particles, LiAlO$_2$ and gehlenite particles as a result of froth flotation tests. These functions provide insight into the combined influence of particle size $d_A$ and shape with aspect ratio $\varphi$ on the separation behavior of different particle systems.

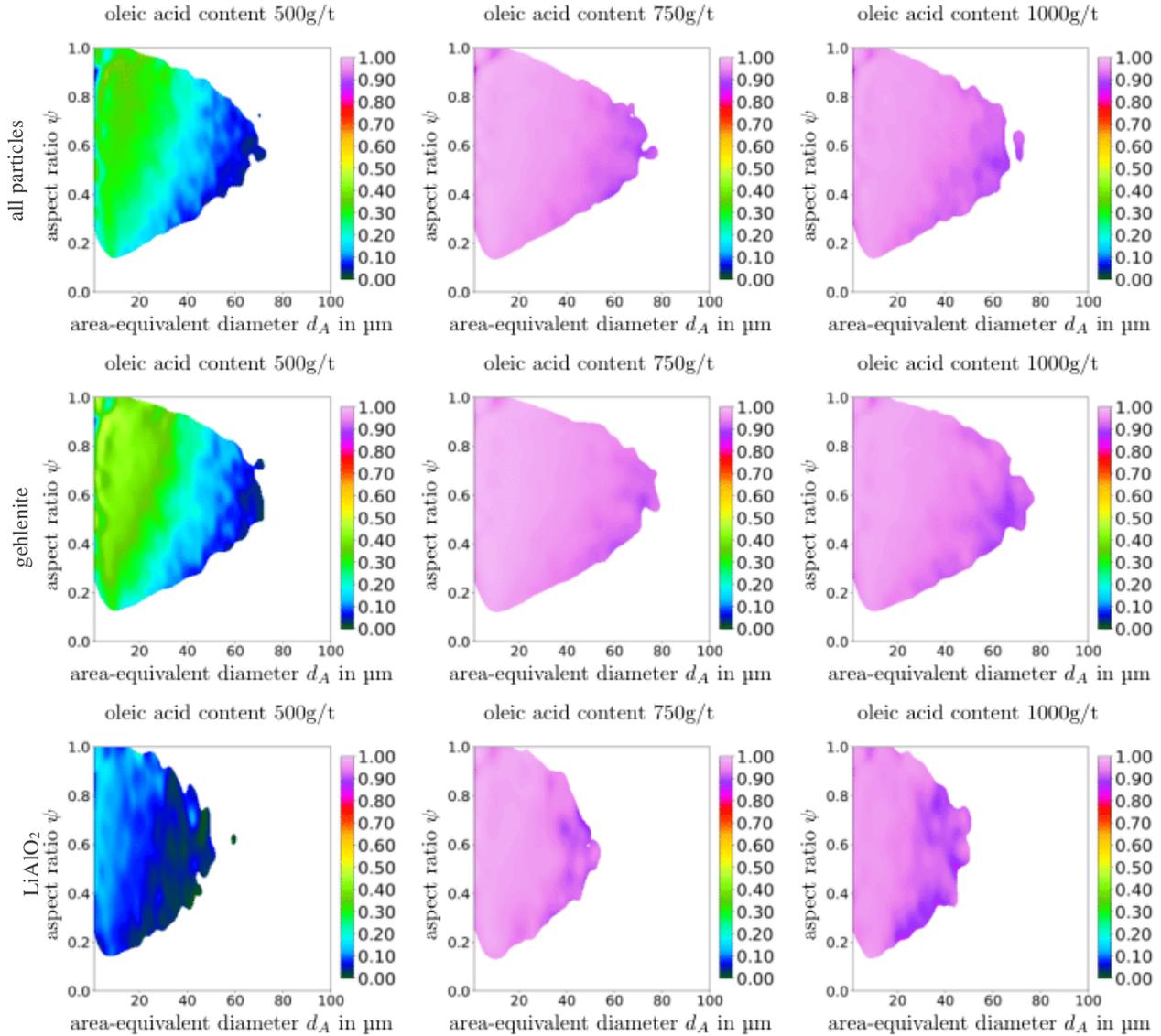

**Figure 7: Bivariate Tromp functions for different particle systems (LiAlO$_2$ (top), gehlenite (middle) and all particles (bottom)), where the Tromp functions are obtained for the different froth flotation tests with 500 g/t OA (left), 750 g/t OA (middle) and 1000 g/t OA (right). The value $T(d_A, \varphi)$ indicates the separation probability of a particles with a certain aspect ratio and area-equivalent diameter to be separated into the concentrate. The white regions correspond to particle descriptor vectors $(d_A, \varphi)$ with $f_n^{\text{feed}}(d_A, \varphi) < 0.001$, i.e., particles with such descriptor vectors are observed with sufficiently high frequency in the feed.**

From the bivariate Tromp functions in Figure 7, it can be seen that a change in OA concentration influences the resulting separation probabilities for all particle systems. For low OA concentration, the separation behaviour of particles in all three particle systems is primarily influenced by particle size. However, particles with an area-equivalent diameter larger than $10 \mu m$ exhibit a higher likelihood of being enriched in the concentrate if they also possess a larger aspect ratio, suggesting that spherical particles are more prone to be separated into the concentrate. For increasing OA concentration almost all particles are separated into the concentrate independently of their size, shape and affiliation to one of the particle systems. But for OA concentration of 1000 g/t the separation probability of larger particles gets lower in comparison to an OA concentration of 750 g/t indicating that a larger concentration does not lead to larger particle enrichment.

## CONCLUSION

This study investigates the flotation of lithium-bearing slags in a Partidge-Smith cell using OA as collector. In general, the assumption that OA as a known collector for spodumene flotation could suit as a selective collector for the upgrading of EnAM $LiAlO_2$ from Li-bearing slags could be not confirmed. In contrast, froth flotation of Li-slag material with a collector concentration of 500 g/t OA resulted in upgrading of the gangue mineral gehlenite. At 750 g/t and 1000 g/t OA almost 90 wt% of the feed mass is recovered in the concentrate, caused by a simultaneously increasing mass and water pull, resulting in low selectivity.

Bivariate Tromp functions, computed from MLA data to analyze the combined influence of particle shape and size, show that the aspect ratio has little influence on the separation, instead, it seems to be driven by the particle size. However, this effect is less significant for higher OA concentrations, as the recovery probability is close to 100 % for almost all particles, regardless of the considered phase.

Future studies should also look into shape factors, other than aspect ratio, such as roundness, as they might be more significant for the flotation behavior. Furthermore, the results show that there is a need to find selective collectors for lithium flotation, as the use of OA did not result in a significant upgrading of $LiAlO_2$.

## FUNDING

This research is partially funded by the German Research Foundation (DFG) via the research projects RU 2184/2-1 and SCHM 997/45-1 within the priority program SPP 2315 "Engineered artificial minerals (EnAM) - A geo-metallurgical tool to recycle critical elements from waste streams".

## ACKNOWLEDGEMENTS

Josephine Roth is gratefully acknowledged to perform the froth flotation experiments. Michael Stoll and Roland Würkert are gratefully acknowledged for the preparation of 12 epoxy embedded samples with an especially developed technique for the MLA measurements.